\documentclass[10pt,final,conference,oneside,twocolumn]{IEEEtran}

\usepackage{graphicx}
\usepackage{amsthm}
\usepackage{mathtools}
\usepackage{bm}
\usepackage{color}

\usepackage{cite}

\ifCLASSINFOpdf
\else
\fi
\usepackage{amsmath,amssymb,amsfonts}
\makeatletter
\def\endthebibliography{%
  \def\@noitemerr{\@latex@warning{Empty `thebibliography' environment}}%
  \endlist
}
\makeatother

\begin{document}
%
\title{Intelligent Reflecting Surface-Aided Backscatter Communications}
%

\author{
	\IEEEauthorblockN{Xiaolun Jia\IEEEauthorrefmark{1}, Jun Zhao\IEEEauthorrefmark{2}, Xiangyun Zhou\IEEEauthorrefmark{1} and Dusit Niyato\IEEEauthorrefmark{2}}
	\IEEEauthorblockA{\IEEEauthorrefmark{1}Research School of Electrical, Energy and Materials Engineering, The Australian National University, Canberra, Australia}
	\IEEEauthorblockA{\IEEEauthorrefmark{2}School of Computer Science and Engineering, Nanyang Technological University, Singapore}
	\IEEEauthorblockA{Email: xiaolun.jia@anu.edu.au, junzhao@ntu.edu.sg, xiangyun.zhou@anu.edu.au, dniyato@ntu.edu.sg}
}

\maketitle

\begin{abstract}
We introduce a novel system setup where a backscatter device operates in the presence of an intelligent reflecting surface (IRS). In particular, we study the bistatic backscatter communication (BackCom) system assisted by an IRS. The phase shifts at the IRS are optimized jointly with the transmit beamforming vector of the carrier emitter to minimize the transmit power consumption at the carrier emitter whilst guaranteeing a required BackCom performance. The unique channel characteristics arising from multiple reflections at the IRS render the optimization problem highly non-convex. Therefore, we jointly utilize the minorization-maximization algorithm and the semidefinite relaxation technique to present an approximate solution for the optimal IRS phase shift design. We also extend our analytical results to the monostatic BackCom system. Numerical results indicate that the introduction of the IRS brings about considerable reductions in transmit power, even with moderate IRS sizes, which can be translated to range increases over the non-IRS-assisted BackCom system.
\end{abstract}


%
\IEEEpeerreviewmaketitle

\section{Introduction}
%
%
%
%

Backscatter communication, or BackCom, has received increasing research interest in recent times, as a potential solution to address the energy efficiency and sustainability of sensor networks under the Internet of Things (IoT). Conventional applications of BackCom include radiofrequency identification (RFID) systems, where RFID tags transmit small data packets to a reader by performing modulation on top of an existing signal. The concept of modulation by reflection has since emerged as a key technology for industrial IoT and pervasive wide-area networking; as a result, the bistatic \cite{Kim14} and ambient \cite{Liu13} architectures have since been proposed, to improve the range of BackCom and its compatibility with radiofrequency (RF) signals that are already modulated. 

Despite the extensive literature on improving BackCom system performance in terms of reliability and throughput, its reliance on external RF signals is still a prominent barrier preventing its widespread deployment. Specifically, the signal power received from BackCom devices, or tags, in monostatic systems, scales inversely with the fourth power of the tag-reader distance. Bistatic systems require dedicated carrier emitters (CEs) placed close to tags in low-interference environments to achieve longer range; and ambient systems suffer from direct-link interference, which incurs high complexity to mitigate, thereby limiting its range to a few meters.

Intelligent reflecting surfaces (IRS) have been recently proposed as a way to modify the wireless propagation medium. An IRS consists of a number of adjustable reflecting elements in either reflectarray or metasurface configurations, and interact directly with impinging signals to alter their amplitudes and phases. In addition to its energy-efficient operation, the coordinated design of phase shifts for a large number of IRS reflectors allows reflected signals to be received constructively (or destructively) at a receiving node \cite{Lia18}. This allows for favorable SNR scaling at the receiver where signals are constructively received, with the SNR shown to scale with the square of the IRS surface area \cite{Ozd19}.

As a result, many works have examined the performance improvements of introducing IRS to a range of communication systems. Work in \cite{Bas19, Han19} studied fundamental metrics of IRS-assisted systems, such as error performance and capacity; while detailed analysis of propagation and path loss in IRS-reflected links was presented in \cite{Ozd19, Ell19}. The joint optimization of IRS phase shifts and transmit beamforming vectors was examined in works such as \cite{Wu19, Pan19}, among many others. More recently, research attention has shifted towards facilitating joint energy and information transfer using IRS through works such as \cite{Wu19b, Tang19b}. However, the use of IRS to support passive communication technologies, specifically BackCom, has received relatively little attention in the literature.

In this paper, we study a system where a BackCom device operates in the presence of a nearby IRS. Given the rapid uptake in IRS research and its expected widespread deployment, it is necessary to consider the performance of existing communication systems, with BackCom being an example, where an IRS is likely to be nearby. To the best of our knowledge, there are few recent works jointly considering IRS and detached BackCom devices, one of which is \cite{Zhao20}. However, \cite{Zhao20} studies the error performance of a non-conventional monostatic system with no direct reader-tag link; whereas no works have considered the standard monostatic or bistatic BackCom architectures assisted by IRS.

The main contributions of this paper are as follows:
\begin{itemize}
\item We introduce an IRS-aided BackCom system where 
the backscatter communication from the tag to the reader is assisted by the IRS. This new BackCom system possesses unique characteristics where signals may be reflected multiple times by the IRS.
\item Specifically considering a bistatic BackCom system where a tag's backscatter communication to a reader is powered by a multi-antenna CE, the IRS reflects the signals from both the CE and the tag. This presents a highly non-trivial design problem on the IRS phase shifts. We jointly optimize the IRS phase shifts and the transmit beamforming at the CE in order to minimize the required transmit power consumption at the CE.
\item We further extend our analysis to a monostatic BackCom system, and obtain the optimal phase shifts for the IRS.
\item Numerical results reveal notable reductions in the required transmit power compared to monostatic and bistatic BackCom systems without IRS, which can be translated to improvements in the link budget and range.
\end{itemize}

\underline{\textit{Notations:}} $j = \sqrt{-1}$ denotes the complex unit, and $\mathbb{R}$ and $\mathbb{C}$ denote the set of real and complex numbers, respectively. $\left| \cdot \right|$ and $\mathrm{Re}\{\cdot\}$ denote the magnitude and the real part of a complex number, respectively. $\mathcal{CN}(\mu, \sigma^{2})$ represents a complex Gaussian distribution with mean $\mu$ and variance $\sigma^{2}$. Vector and matrix quantities are denoted using lowercase and uppercase boldface letters, respectively, as in $\bm{a}$ and $\bm{A}$. $\bm{I}$ denotes the identity matrix of variable size. $\left\lVert \bm{a} \right\rVert$ denotes the Euclidean norm of a vector; and $\mathrm{tr}(\bm{A})$ and $\bm{A}^{H}$ denote the trace and the Hermitian transpose of $\bm{A}$, respectively.


\section{System and Signal Model}

\subsection{System Setup}

\begin{figure}[!t]
\centering
\includegraphics[width=0.44\textwidth]{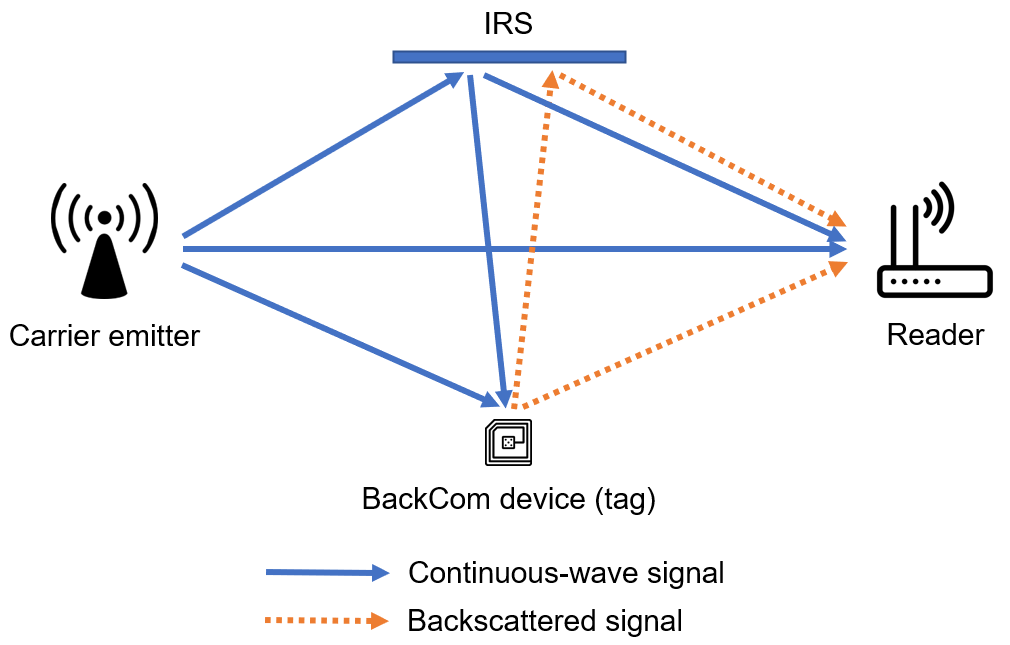}
\caption{System setup for the IRS-assisted bistatic BackCom system.}
\label{fig_1}
\end{figure}

We consider an IRS-aided bistatic BackCom system with an $L$-antenna CE, a single-antenna tag, a single-antenna reader, and an IRS with $N$ reflecting elements. Hereafter, we denote the CE, tag, IRS and reader by $C$, $T$, $I$ and $R$, respectively. A system diagram is shown in Fig. \ref{fig_1}. Despite the simplicity of the system, the design complexity arises from the fact that the IRS must balance the signal reflections from both the CE-to-tag and tag-to-reader links. The associated phase shift design problem is highly nonconvex, and is explored in Section III.

The CE transmits a continuous-wave signal with power $P$ and beamforming vector $\bm{w} \in \mathbb{C}^{L \times 1}$ to power the tag's communication, where $P = \left\lVert \bm{w} \right\rVert^{2}$.

The tag has $J$ load impedances connected to its antenna. We assume that the tag has an off-state where the load and antenna impedances are perfectly matched, resulting in no reflection and no data transmission. We consider a generalized tag configuration where the tag could be either passive or semipassive, with circuit power consumption $\xi$. Where $\xi = 0$, the tag is powered by an on-board battery; otherwise, a portion of energy from the incoming signal is used to power the tag.

Each reflecting element of the IRS is modeled as a diffuse reflector \cite{Ozd19}. As diffuse reflections incur significant path loss, we ignore signals which undergo two or more reflections at the tag. However, since the IRS needs to balance the signals in both the $C$-$I$-$T$ and $T$-$I$-$R$ links, this assumption does not apply to signals reflected exactly two times by the IRS. That is, the $C$-$I$-$T$-$I$-$R$ link is still considered.

We consider the ideal assumption that the reader has perfect knowledge of the channel state information (CSI) of all channels. This assumption allows us to characterize the upper bound on the system performance. The evaluation of any channel estimation methods for an IRS-aided BackCom system is outside the scope of this work. However, we point out that works such as \cite{Mis19a, Lin19, Yang19} present methods to assist with estimation of channels involving the CE, tag and reader; whereas individual tag-IRS reflector channels may be estimated by switching off the reflectors one at a time.

\subsection{Signal Model}

The tag modulates its data onto an incident signal by switching between its impedances \cite{Kim14, Liu13}. Its baseband signal is denoted by $b(t)$, which is the tag's time-varying reflection coefficient.\footnote{Realistically, the tag baseband signal is of the form $b(t) = A - \Gamma(t)$, where $\Gamma(t)$ is the reflection coefficient, and $A$ is the tag's antenna structural mode, and determines the default amount of signal reflection in the off-state. However, $A$ is a constant, and can hence be subtracted from the received signal in post-processing. Therefore, we do not consider it here.} The reflection coefficient takes on values $b(t) \in \{b_{1}, \ldots, b_{J}\}$, where each value corresponds to an impedance; moreover, $|b_{i}| = |b_{j}|, \ \forall i, j$, and $|b_{i}| \leq 1, \ \forall i$. The power splitting coefficient at the tag is denoted by $\alpha \in [0, 1]$, where $\alpha$ denotes the proportion of the signal to be reflected, with $1\!-\!\alpha$ proportion of the signal used to power the circuit.

The continuous-wave signal transmitted by the CE is given by $s(t)$. Denote the channels from the CE to tag, CE to IRS, CE to reader, IRS to tag, IRS to reader and tag to reader by $\bm{h}_{CT} \in \mathbb{C}^{1 \times L}$, $\bm{H}_{CI} \in \mathbb{C}^{N \times L}$, $\bm{h}_{CR} \in \mathbb{C}^{1 \times L}$, $\bm{h}_{TI}^{H} \in \mathbb{C}^{1 \times N}$, $\bm{h}_{RI}^{H} \in \mathbb{C}^{1 \times N}$ and $h_{TR} \in \mathbb{C}^{1 \times 1}$, respectively. Each IRS element $n \in \{1, \ldots, N\}$ reflects the sum of all incident signals with a phase shift, denoted by $\theta_{n}$. We assume the amplitude gain of each IRS element to be unity. Let the vector containing the phase shifts of all elements be $\bm{\theta} = \left[ \theta_{1}, \ldots, \theta_{N} \right]^{T}$, where $\theta_{n} \in [0, 2\pi), \ \forall n$. The IRS phase shift matrix is then given by $\bm{\Theta} = \mathrm{diag} \left( e^{j \theta_{1}}, \ldots, e^{j \theta_{N}} \right)$.

Linear transmit precoding is assumed at the CE with a single beamforming vector $\bm{w}$. The transmitted signal is then $\bm{x}_{C} = \bm{w} s(t)$. The signal received at the tag consists of the direct $C$-$T$ link and the reflected $C$-$I$-$T$ link, and is given by
\begin{equation}
y_{T} = \left( \bm{h}_{TI}^{H} \bm{\Theta} \bm{H}_{CI} + \bm{h}_{CT} \right) \bm{w} s(t).   \label{tagReceivedSignal}
\end{equation}
No noise term is added at the tag, as no signal processing is performed. The part of the signal reflected by the tag is
\begin{equation}
x_{T, r} = \sqrt{\alpha} b(t) y_{T}.    \label{tagReflectedSignal}
\end{equation}
The remainder of the signal is used to power the circuit, whose squared magnitude, denoted by $E_{h}$, can be given by
\begin{equation}
E_{h} = \eta (1 - \alpha) |y_{T}|^{2},    \label{tagHarvestedSignal}
\end{equation}
where $\eta \in [0, 1]$ is the energy harvesting efficiency. As a result, the circuit constraint is given by
\begin{equation}
\eta (1 - \alpha) \left| \left( \bm{h}_{TI}^{H} \bm{\Theta} \bm{H}_{CI} + \bm{h}_{CT} \right) \bm{w} \right|^{2} \geq \xi, \label{circuitConstraint}
\end{equation}
with $\xi$ being the circuit power consumption in Watts. The signal received at the reader consists of those arrived directly from the CE, backscattered from the tag, and reflected from the IRS. After removing the constant (unmodulated) continuous-wave signals in the $C$-$R$ and $C$-$I$-$R$ links, the signal to be processed can be written as
\begin{multline}
y_{R} = \sqrt{\alpha} b(t) \left( \bm{h}_{RI}^{H} \bm{\Theta} \bm{h}_{TI} + h_{TR} \right) \\ \times \left( \bm{h}_{TI}^{H} \bm{\Theta} \bm{H}_{CI} + \bm{h}_{CT} \right) \bm{w} s(t) + n_{R},    \label{readerReceivedSignalShort}
\end{multline}
where $n_{R} \sim \mathcal{CN}(0, \sigma_{R}^{2})$ is the noise at the reader. The signal-to-noise ratio (SNR) is thus defined as the squared magnitude of the noiseless part of (\ref{readerReceivedSignalShort}) divided by $\sigma_{R}^{2}$, using the received signal power when the tag is in its off-state as reference.


\section{Transmit Power Minimization Problem}

In this paper, we study the transmit power minimization problem at the CE, subject to an SNR constraint on the tag's information transmission to the reader. The transmit power minimization problem is appealing for a BackCom system, as it allows not only the determination of optimal IRS and BackCom device parameters, but allows the possibility of translating the power savings from the obtained solution to an increase in the power budget and hence range.

To solve the problem, we jointly optimize the transmit beamforming vector at the CE, the phase shift coefficients at the IRS, and the power splitting coefficient at the tag. We begin by presenting the problem for a bistatic BackCom system, and extend it to the monostatic architecture in the sequel. The problem can be written as (Problem $\mathbf{P}$):
\begin{subequations}
\begin{alignat}{3}
&\!\min_{\bm{w}, \bm{\theta}, \alpha} & \qquad & \left\lVert \bm{w} \right\rVert^{2} \label{PA} \\
& \text{s.t.} & & \alpha |b(t)|^{2} \left| \left( \bm{h}_{RI}^{H} \bm{\Theta} \bm{h}_{TI} + h_{TR} \right) \right. \nonumber \\
& & & \qquad \times \left. \left( \bm{h}_{TI}^{H} \bm{\Theta} \bm{H}_{CI} + \bm{h}_{CT} \right) \bm{w} \right|^{2} \geq \gamma_{th} \sigma_{R}^{2}, \label{PB} \\
& & & \eta (1 - \alpha) \left| \left( \bm{h}_{TI}^{H} \bm{\Theta} \bm{H}_{CI} + \bm{h}_{CT} \right) \bm{w} \right|^{2} \geq \xi, \label{PC} \\
& & & 0 \leq \alpha \leq 1, \label{PD} \\
& & & 0 \leq \theta_{n} \leq 2 \pi,  \forall n \in \{1, \ldots, N\}, \label{PE}
\end{alignat}
\end{subequations}
where (\ref{PB}) is the tag's SNR constraint, (\ref{PC}) is the tag's circuit power constraint, (\ref{PD}) is the splitting coefficient constraint, and (\ref{PE}) is the phase shift constraint for each IRS element.

For conciseness, the following substitutions are made:
\begin{align}
\bm{H}_{1}(\bm{\Theta}) &\triangleq \bm{h}_{TI}^{H} \bm{\Theta} \bm{H}_{CI} + \bm{h}_{CT}, 	\label{1H} \\
\bm{H}_{2}(\bm{\Theta}) &\triangleq \bm{h}_{RI}^{H} \bm{\Theta} \bm{h}_{TI} + h_{TR}.   \label{2H}
\end{align}
Equations (\ref{1H}) and (\ref{2H}) correspond to the combined channel gains for the CE-to-tag and tag-to-reader links, respectively, which include the reflected signal paths from the IRS.

\subsection{IRS Phase Shift Design: No Circuit Power Constraint}

We begin by noting that when there is one semipassive tag (i.e., $\xi = 0$), optimal beamforming can be achieved using maximum ratio transmission (MRT). Therefore, the optimal $\bm{w}$ in Problem $\mathbf{P}$ is simply
\begin{equation}
\bm{w}^{*}\!=\!\sqrt{P} \frac{\left[ \bm{H}_{2}(\bm{\Theta}) \bm{H}_{1}(\bm{\Theta}) \right]^{H}}{\left\lVert  \bm{H}_{2}(\bm{\Theta}) \bm{H}_{1}(\bm{\Theta}) \right\rVert}.     \label{optimalBV}
\end{equation}
Substituting (\ref{optimalBV}) into Problem $\mathbf{P}$ (without constraint (\ref{PC})), we can rewrite it directly in terms of $P$ (Problem $\mathbf{P1}$):
\begin{subequations}
\begin{alignat}{3}
&\!\min_{P, \bm{\theta}, \alpha} & \qquad & P \label{P1A} \\
& \text{s.t.} & & P \alpha |b(t)|^{2} \left\lVert \bm{H}_{2}(\bm{\Theta}) \bm{H}_{1}(\bm{\Theta}) \right\rVert^{2} \geq \gamma_{th} \sigma_{R}^{2}, \label{P1B} \\
& & & 0 \leq \theta_{n} \leq 2 \pi,  \forall n \in \{1, \ldots, N\}. \label{P1D}
\end{alignat}
\end{subequations}

To numerically quantify $P^{*}$, the optimal IRS phase shifts $\bm{\Theta}$ need to be determined. Rearranging (\ref{P1B}) to solve for $P$, we can directly maximize the denominator of the resulting expression over $\bm{\Theta}$ as follows (Problem $\mathbf{P2}$-$\mathbf{nc}$):
\begin{subequations}
\begin{alignat}{3}
&\!\max_{\bm{\theta}} & \qquad & |\bm{H}_{2}(\bm{\Theta})|^{2} \left\lVert \bm{H}_{1}(\bm{\Theta}) \right\rVert^{2},  \label{P2nc} \\
& \text{s.t.} & & 0 \leq \theta_{n} \leq 2 \pi,  \forall n \in \{1, \ldots, N\}. \label{revisedProblem4}
\end{alignat}
\end{subequations}
First, $\left\lVert \bm{H}_{1}(\bm{\Theta}) \right\rVert^{2}$ can be rewritten as follows:
\begin{multline}
\left\lVert \bm{H}_{1}(\bm{\Theta}) \right\rVert^{2} = \bm{v}^{H} \bm{\Phi}_{CIT} \bm{\Phi}_{CIT}^{H} \bm{v} + \bm{v}^{H} \bm{\Phi}_{CIT} \bm{h}_{CT}^{H} \\
 + \bm{h}_{CT} \bm{\Phi}_{CIT}^{H} \bm{v} + \left\lVert \bm{h}_{CT} \right\rVert^{2},   \label{QCQPStandard}
\end{multline}
where $\bm{\Phi}_{CIT} \triangleq \textrm{diag}(\bm{h}_{TI}^{H}) \bm{H}_{CI}$ and $\bm{v} \triangleq \left[ e^{j \theta_{1}}, \ldots, e^{j \theta_{N}} \right]^{H}$, with $|v_{n}|^{2} = 1, \forall n$. It is evident that (\ref{QCQPStandard}) is of quadratic form, and can therefore be rewritten in matrix form as
\begin{equation}
\left\lVert \bm{H}_{1}(\bm{\Theta}) \right\rVert^{2} = \bm{\bar{v}}^{H} \bm{R} \bm{\bar{v}} + \lVert \bm{h}_{CT} \rVert^{2}, \label{expanded1stHop}
\end{equation}
with
\begin{equation}
\bm{R} = 
\begin{bmatrix}
\bm{\Phi}_{CIT} \bm{\Phi}_{CIT}^{H} & \bm{\Phi}_{CIT} \bm{h}_{CT}^{H} \\
\bm{h}_{CT} \bm{\Phi}_{CIT}^{H} & 0
\end{bmatrix},
\hspace{5mm} \bm{\bar{v}} = 
\begin{bmatrix}
\bm{v} \\
1
\end{bmatrix}.  \label{R}
\end{equation}
The expanded form of $|\bm{H}_{2}(\bm{\Theta})|^{2}$ can be given by
\begin{multline}
|\bm{H}_{2}(\bm{\Theta})|^{2} = \bm{v}^{H} \bm{\Phi}_{TIR} \bm{\Phi}_{TIR}^{H} \bm{v} + \bm{v}^{H} \bm{\Phi}_{TIR} h_{TR}^{H} \\
 + h_{TR} \bm{\Phi}_{TIR}^{H} \bm{v} + |h_{TR}|^{2},    \label{expandedAlpha}
\end{multline}
with $\bm{\Phi}_{TIR} \triangleq \textrm{diag}(\bm{h}_{RI}^{H}) \bm{h}_{TI}^{H}$. Equation (\ref{expandedAlpha}) can also be re-written in matrix form as follows:
\begin{equation}
|\bm{H}_{2}(\bm{\Theta})|^{2} = \bm{\bar{v}}^{H} \bm{S} \bm{\bar{v}} + |h_{TR}|^{2}, \label{QCQP2}
\end{equation}
with
\begin{equation}
\bm{S} = 
\begin{bmatrix}
\bm{\Phi}_{TIR} \bm{\Phi}_{TIR}^{H} & \bm{\Phi}_{TIR} h_{TR}^{H} \\
h_{TR} \bm{\Phi}_{TIR}^{H} & 0
\end{bmatrix}.  \label{S}
\end{equation}
With the additional $|\bm{H}_{2}(\bm{\Theta})|^{2}$ term, it is clear that the problem under the BackCom system is considerably different to those in existing works, which tend to optimize quadratic objective functions. As a result, denoting the original objective function in (\ref{P2nc}) by $F$, the product of (\ref{expanded1stHop}) and (\ref{QCQP2}) is given by
\begin{align}
F(\bm{\bar{v}}) &= (\bm{\bar{v}}^{H} \bm{R} \bm{\bar{v}} + c_{1}) (\bm{\bar{v}}^{H} \bm{S} \bm{\bar{v}} + c_{2}) \nonumber \\
    &= \bm{\bar{v}}^{H} \bm{S} \bar{\bm{v}} \bar{\bm{v}}^{H} \bm{R} \bm{\bar{v}}\!+\!c_{1} \bm{\bar{v}}^{H} \bm{S} \bm{\bar{v}}\!+\!c_{2} \bm{\bar{v}}^{H} \bm{R} \bm{\bar{v}}\!+\!c_{1} c_{2}, \label{revisedOF}
\end{align}
with $c_{1} = \lVert \bm{h}_{CT} \rVert^{2}$ and $c_{2} = |h_{TR}|^{2}$. Equation (\ref{revisedOF}) is a quartic polynomial in $\bar{\bm{v}}$. Normally, to optimize a quadratic form such as (\ref{expanded1stHop}), we can let $\bm{V} \triangleq \bm{\bar{v}} \bm{\bar{v}}^{H}$, and use the identity $\bar{\bm{v}}^{H} \bm{R} \bar{\bm{v}} = \mathrm{tr}(\bm{R} \bar{\bm{v}} \bar{\bm{v}}^{H})$ to recast (\ref{expanded1stHop}) as a function of $\bm{V}$, which is rank-one. However, here we cannot invoke the trace identity on (\ref{revisedOF}), as the first resulting trace term, $\mathrm{tr}(\bm{S} \bm{V} \bm{R} \bm{V})$, is generally nonconvex. It has also been noted in \cite{Luo10} that it is NP-hard to optimize (minimize) polynomials of degree $4$, meaning that a closed-form, optimal solution to Problem $\mathbf{P2}$-$\mathbf{nc}$ is generally not available. 

To address this challenging issue, we use the semidefinite relaxation (SDR) technique nested within the minorization-maximization (MM) algorithm. In each iteration, we find a convex minorizing function to (\ref{revisedOF}) and obtain a relaxed solution to an equivalent of Problem $\mathbf{P2}$-$\mathbf{nc}$ (outlined below) via SDR. Then, the solution from randomization is used to refine the minorizer to be closer to the maximum of (\ref{revisedOF}). The process is repeated until convergence of the MM algorithm.

We construct a minorizer to a function $f(\bm{x}): \mathbb{C}^{N}\!\rightarrow\!\mathbb{R}$ with bounded curvature, such as the absolute value of a complex-valued function, by modifying \cite[Lemma 12]{Sun17} (which applies to functions that are $\mathbb{R}^{N} \rightarrow \mathbb{R}$) to take the second-order Taylor expansion:
\begin{equation}
f(\bm{x})\!\geq\!f(\bm{x}_{0})\!+\!\mathrm{Re}\!\left\{ \nabla f(\bm{x}_{0})^{H} (\bm{x}\!-\!\bm{x}_{0}) \right\}\!-\!\frac{\ell}{2} \left\lVert \bm{x}\!-\!\bm{x}_{0} \right\rVert^{2},   \label{minorizer}
\end{equation}
where $\bm{x}_{0} \in \mathbb{C}^{N}$ is any point, and $\ell$ is the maximum curvature of $f(\bm{x})$. Applying (\ref{minorizer}) to (\ref{revisedOF}), we obtain
\begin{align}
F(\bar{\bm{v}}) &\geq \bar{\bm{v}}_{0}^{H} \bm{S} \bar{\bm{v}}_{0} \bar{\bm{v}}_{0}^{H} \bm{R} \bar{\bm{v}}_{0} + c_{1} \bar{\bm{v}}_{0}^{H} \bm{S} \bar{\bm{v}}_{0} + c_{2} \bar{\bm{v}}_{0}^{H} \bm{R} \bar{\bm{v}}_{0} + c_{1} c_{2} \nonumber \\
    & \qquad + \bar{\bm{v}}_{0}^{H} \bm{T} (\bar{\bm{v}} - \bar{\bm{v}}_{0}) + (\bar{\bm{v}} - \bar{\bm{v}}_{0})^{H} \bm{T} \bar{\bm{v}}_{0} \nonumber \\
    & \qquad - \frac{\ell}{2}(\bar{\bm{v}}^{H} \bar{\bm{v}} - \bar{\bm{v}}^{H} \bar{\bm{v}}_{0} - \bar{\bm{v}}_{0}^{H} \bar{\bm{v}} + \left\lVert \bar{\bm{v}}_{0} \right\rVert^{2}) \nonumber \\
    &= -\frac{\ell}{2}(\bar{\bm{v}}^{H} \bar{\bm{v}} - \bar{\bm{v}}^{H} \bar{\bm{v}}_{0} - \bar{\bm{v}}_{0}^{H} \bar{\bm{v}} + \left\lVert \bar{\bm{v}}_{0} \right\rVert^{2}) \nonumber \\
    & \qquad + \bar{\bm{v}}_{0}^{H} \bm{T} \bar{\bm{v}} + \bar{\bm{v}}^{H} \bm{T} \bar{\bm{v}}_{0} + c \nonumber \\
    &= -\frac{\ell}{2} \left( \bar{\bm{v}}^{H} \bm{I} \bar{\bm{v}} + \bar{\bm{v}}^{H} \left( -\frac{2}{\ell} \bm{T} \bar{\bm{v}}_{0} - \bm{I} \bar{\bm{v}}_{0} \right) \right. \nonumber \\
    & \qquad \left. + \left( -\frac{2}{\ell} \bm{T} \bar{\bm{v}}_{0} - \bm{I} \bar{\bm{v}}_{0} \right)^{H} \bm{\bar{v}} \right) + c,   \label{quadraticApprox}
\end{align}
where $\bm{T} \triangleq \bm{R} \bar{\bm{v}}_{0} \bar{\bm{v}}_{0}^{H} \bm{S} + \bm{S} \bar{\bm{v}}_{0} \bar{\bm{v}}_{0}^{H} \bm{R} + c_{2} \bm{R} + c_{1} \bm{S}$, which is a Hermitian matrix; and $c$ denotes the cumulative sum of all constant terms and terms involving only $\bar{\bm{v}}_{0}$. The right-hand side of (\ref{quadraticApprox}) is of quadratic form, and can be rewritten in matrix form as $\bar{\bar{\bm{v}}}^{H} \bm{U} \bar{\bar{\bm{v}}}$, where
\begin{equation}
\bm{U} = 
-\begin{bmatrix}
\bm{I} & -\frac{2}{\ell} \bm{T} \bar{\bm{v}}_{0} - \bm{I} \bar{\bm{v}}_{0} \\
(-\frac{2}{\ell} \bm{T} \bar{\bm{v}}_{0} - \bm{I} \bar{\bm{v}}_{0})^{H} & 0
\end{bmatrix},
\bar{\bar{\bm{v}}} = 
\begin{bmatrix}
\bar{\bm{v}} \\
1
\end{bmatrix}.  \label{U}
\end{equation}
Then, letting $\bar{\bar{\bm{V}}} = \bar{\bar{\bm{v}}} \bar{\bar{\bm{v}}}^{H}$, Problem $\mathbf{P2}$-$\mathbf{nc}$ can now be recast as the following equivalent problem (Problem $\mathbf{P2.1}$-$\mathbf{nc}$)
\begin{subequations}
\begin{alignat}{3}
&\!\max_{\bar{\bar{\bm{V}}}} & \quad & \mathrm{tr}(\bm{U} \bar{\bar{\bm{V}}}) + c, \\
& \text{s.t.} & & \bar{\bar{\bm{V}}}_{n, n} = 1,  \forall n \in \{1, \ldots, N+2\}, \\
& & & \bar{\bar{\bm{V}}} \succeq 0, \\
& & & \mathrm{rank}(\bar{\bar{\bm{V}}}) = 1.  \label{revisedProblem6}
\end{alignat}
\end{subequations}
Dropping the rank-one constraint on $\bar{\bar{\bm{V}}}$, Problem $\mathbf{P2.1}$-$\mathbf{nc}$ becomes a convex semidefinite program (SDP), and can be solved straightforwardly using CVX \cite{cvx}. An approximate rank-one solution $\bar{\bar{\bm{V}}}_{SDR}$ can then be obtained using Gaussian randomization.  The decomposed vector $\bar{\bar{\bm{v}}}_{SDR}$ is then substituted into (\ref{quadraticApprox}) as $\bar{\bm{v}}_{0}$ to obtain a new $\bm{U}$, and the process is repeated until convergence. Note that $F(\bar{\bm{v}})$ in (\ref{revisedOF}) is bounded above, as $\bm{R}$ and $\bm{S}$ are constant matrices, and $\left\lVert \bar{\bm{v}} \right\rVert^{2} = N$, which is a finite constant. Therefore, the MM algorithm is guaranteed to converge to at least a local optimum.

\subsection{IRS Phase Shift Design: Finite Circuit Power Constraint}

In this subsection, we propose an algorithm to determine $P^{*}$ and $\bm{\Theta}$ under a nonzero circuit power constraint, i.e., $\xi > 0$. As such, Problem $\mathbf{P1}$ requires constraint (\ref{PC}) to be included, which can be rewritten as
\begin{equation}
P \eta (1 - \alpha) \left\lVert \bm{H}_{1}(\bm{\Theta})  \right\rVert^{2} \geq \xi. 	\label{CPS}
\end{equation}
By inspection in conjunction with constraint (\ref{P1B}) in Problem $\mathbf{P1}$, the minimum transmit power must meet both the SNR and circuit power constraints with equality. From (\ref{P1B}) and (\ref{CPS}), an intermediate expression for $P^{*}$ is given by
\begin{equation}
P'\!=\!\max\!\left\{\!\frac{\gamma_{th} \sigma_{R}^{2}}{\alpha |b(t)|^{2}\!\left\lVert \bm{H}_{2}(\bm{\Theta}) \bm{H}_{1}(\bm{\Theta}) \right\rVert^{2}},\! \frac{\xi}{\eta (1\!-\!\alpha)\!\left\lVert \bm{H}_{1}(\bm{\Theta}) \right\rVert^{2}}\!\right\}\!.  \label{suboptimalPower}
\end{equation}
As $\alpha$ is increased from $0$ to $1$, the first term of (\ref{suboptimalPower}) is monotonically decreasing; while the second term of (\ref{suboptimalPower}) is monotonically increasing. Therefore, the optimal value of $\alpha^{*}$ that minimizes $P$ is found by equating the two terms:
\begin{equation}
\alpha^{*}\!=\!\frac{\eta \gamma_{th} \sigma_{R}^{2} \left\lVert \bm{H}_{1}(\bm{\Theta}) \right\rVert^{2}}{\xi |b(t)|^{2} \left\lVert \bm{H}_{2}(\bm{\Theta}) \bm{H}_{1}(\bm{\Theta}) \right\rVert^{2}\!+\!\eta \gamma_{th} \sigma_{R}^{2} \left\lVert \bm{H}_{1}(\bm{\Theta}) \right\rVert^{2}}. \label{optimalRC}
\end{equation}
The minimum transmit power is then found by substituting $\alpha^{*}$ into either term in (\ref{suboptimalPower}):
\begin{equation}
P^{*} = \frac{\gamma_{th} \sigma_{R}^{2} + \frac{\xi}{\eta} |b(t)|^{2} \left| \bm{H}_{2}(\bm{\Theta}) \right|^{2}}{|b(t)|^{2} \left\lVert \bm{H}_{2}(\bm{\Theta}) \bm{H}_{1}(\bm{\Theta}) \right\rVert^{2}}. \label{optimalPower}
\end{equation}
Note that $\bm{\Theta}$ appears in both the numerator and denominator of (\ref{optimalPower}). Therefore, fractional programming techniques can be used to obtain a solution for $\bm{\Theta}$. The minimization of a fractional objective function $F(\bm{x}) = \frac{A(\bm{x})}{B(\bm{x})}$ over $\bm{x}$ is equivalent to maximizing its reciprocal. Hence, we can use the Dinkelbach transform \cite{Din67} to rewrite the original form as (Problem $\mathbf{P2}$-$\mathbf{c}$):
\begin{equation}
\begin{aligned}
&\!\max_{\bm{\theta}} & & \frac{|b(t)|^{2} \left\lVert \left( \bm{h}_{RI}^{H} \bm{\Theta} \bm{h}_{TI} + h_{TR} \right) \left( \bm{h}_{TI}^{H} \bm{\Theta} \bm{H}_{CI} + \bm{h}_{CT} \right) \right\rVert^{2}}{\gamma_{th} \sigma_{R}^{2} + \frac{\xi}{\eta} |b(t)|^{2} \left| \bm{h}_{RI}^{H} \bm{\Theta} \bm{h}_{TI} + h_{TR} \right|^{2}}. \\
& \text{s.t.} & & 0 \leq \theta_{n} \leq 2 \pi,  \forall n \in \{1, \ldots, N\}. \label{noCircuitConstraint}
\end{aligned}
\end{equation}
Denoting the numerator and denominator of (\ref{noCircuitConstraint}) by $A(\bm{\Theta})$ and $B(\bm{\Theta})$, respectively, Problem $\mathbf{P2}$-$\mathbf{c}$ can then be readily solved by maximizing over $A(\bm{\Theta}) - y B(\bm{\Theta})$, with the quantity $y^{(i+1)} = \frac{A(\bm{\Theta}^{(i)})}{B(\bm{\Theta}^{(i)})}$ updated at every iteration $i$. However, each iteration requires convergence of the MM algorithm outlined in the previous subsection, which incurs significant complexity. Therefore, we propose approximate solutions for the circuit constraint-limited and noise-limited regimes. In the former, where $\frac{\xi}{\eta} |b(t)|^{2} \left| \bm{h}_{RI}^{H} \bm{\Theta} \bm{h}_{TI} + h_{TR} \right|^{2} \gg \gamma_{th} \sigma_{R}^{2}$, the minimized transmit power can be obtained in a similar manner as the single-user alternating optimization algorithm in \cite{Wu19}, where the alternating steps are performed over $\bm{w}$ and $\bm{\Theta}$. In the latter, where $\frac{\xi}{\eta} |b(t)|^{2} \left| \bm{h}_{RI}^{H} \bm{\Theta} \bm{h}_{TI} + h_{TR} \right|^{2} \ll \gamma_{th} \sigma_{R}^{2}$, the problem simplifies to be similar to Problem $\mathbf{P2}$-$\mathbf{nc}$, and can be solved using the proposed method in Section III-A.

\subsection{Extension to Monostatic BackCom Systems}

The analysis of a monostatic IRS-aided BackCom system is a special, simpler case of that for bistatic systems in the two previous subsections. Given a single antenna at the reader, all previous channel gains with subscript $C$ are changed to $R$. Assuming reciprocal channels, fewer unique channels are present ($R$-$T$, $R$-$I$, $T$-$I$ compared to $C$-$T$, $C$-$I$, $T$-$I$, $T$-$R$, $I$-$R$), and leads to the following rearrangement of (\ref{1H}):
\begin{equation}
\bm{H}_{1}(\bm{\Theta}) = \bm{h}_{TI}^{H} \bm{\Theta} \bm{h}_{RI} + h_{TR}^{H} = \bm{H}_{2}(\bm{\Theta})^{H}. 	\label{1HR}
\end{equation}
As a result, the function to maximize becomes
\begin{equation}
F(\bm{\Theta}) = \left\lVert \bm{H}_{2}(\bm{\Theta}) \bm{H}_{2}(\bm{\Theta})^{H} \right\rVert^{2} = \left| \left|\bm{H}_{2}(\bm{\Theta})\right|^{2} \right|^{2}.	\label{monostaticOF}
\end{equation}
Assuming semipassive tags, we can then maximize $\left|\bm{H}_{2}(\bm{\Theta})\right|^{2}$, with the optimal solution for each individual phase shift being \begin{equation}
\theta_{n}^{*} = \theta_{TR} - \theta_{RI,n}^{H} - \theta_{TI,n}, 	\label{optimalMono}
\end{equation}
where $\theta_{n}^{*}$ is the optimal phase of the $n$-th IRS element; and $\theta_{TR}$, $\theta_{RI,n}^{H}$ and $\theta_{TI,n}$ are the phases of the channels from tag to reader, the $n$-th IRS element to the reader, and the tag to the $n$-th IRS element, respectively. Due to the simpler form of the objective function in (\ref{monostaticOF}), the minimized transmit power expression is also simpler, with $P^{*} = \gamma_{th} \sigma_{R}^{2} |\bm{H}_{2}(\bm{\Theta^{*}})|^{-4}$.


\section{Numerical Results}

In this section, we numerically evaluate the transmit power of the CE in the IRS-aided BackCom system. We assume the CE to have $L = 4$ antennas and carrier frequency $915$ MHz. The CE is located at the origin and the reader is located at $[100, 0]$, with all coordinates being in meters hereafter. We adopt the path loss model in \cite{Ell19}, which applies to both near- and far-field IRS transmissions.\footnote{The path losses for the $C$-$I$, $I$-$T$ and $I$-$R$ links are individually computed based on \cite[Eq. (21)-(22)]{Ell19} and absorbed into $\bm{H}_{CI}$, $\bm{h}_{TI}^{H}$ and $\bm{h}_{RI}^{H}$, respectively; while the path losses for the $C$-$T$ and $T$-$R$ links are absorbed into $\bm{h}_{CT}$ and $h_{TR}$, respectively, consistent with \cite{Ozd19}.} The IRS is oriented towards direction $[0 \ $-$1]^{T}$, and is equipped with reflecting elements of width $\lambda$ without loss of generality \cite{Ozd19}. We consider an outdoor scenario with path loss exponent $2.1$; for simplicity of analysis, the channel coefficients have unit magnitude with random phases drawn from a uniform distribution over $[0, 2 \pi)$. Unless otherwise noted, the number of channel realizations is $100$; the tag is assumed to be semi-passive with $\xi = 0$ and $\eta = 1$; $|b(t)| = 1$ for all tag impedances; the SNR requirement at the reader is $\gamma_{th} = 8$ dB; and the noise power at the reader is $\sigma_{R}^{2} = -110$ dBm. The curvature upper bound for equation (\ref{optimalRC}) is set to $\ell = 2.5 \times 10^{-16}$ and the convergence threshold for the objective function involving $\bm{\Theta}$ is $10^{-4}$.

\begin{figure}[!t]
\centering
\includegraphics[width=0.485\textwidth]{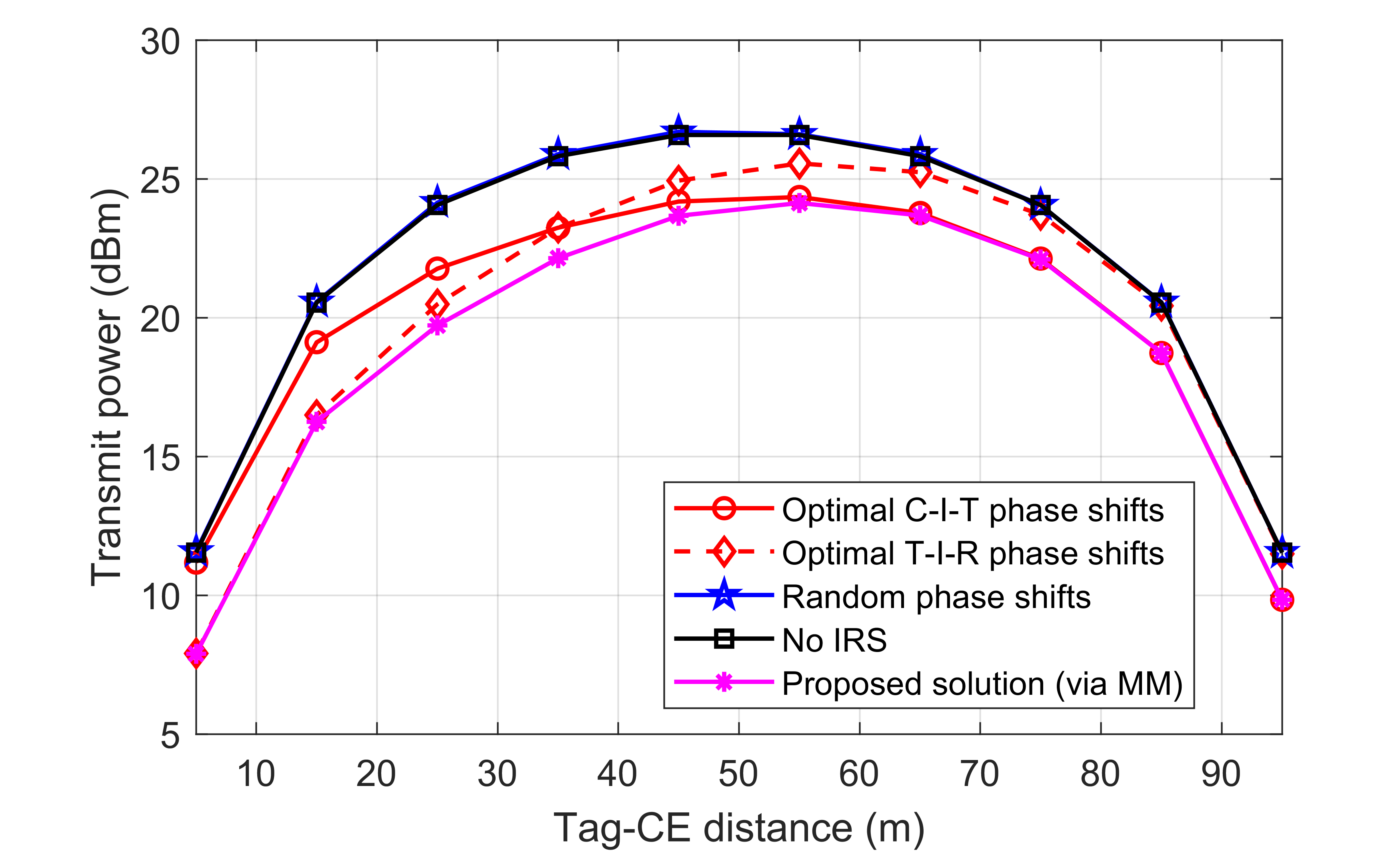}
\caption{Minimized CE transmit power vs. tag location.}
\label{fig_2}
\end{figure}

Fig.~2 shows the minimized transmit power of the CE as the tag is moved on a straight line between the CE and the reader. For this experiment, $N = 64$; the IRS is located at $[20, 20]$; and the tag location is between $[5, 0]$ and $[95, 0]$. In addition to solving Problem $\mathbf{P2.1}$-$\mathbf{nc}$ using the algorithm outlined in Section III-A, we compute the suboptimal transmit power using several benchmark schemes, including: a no-IRS system; random IRS phase shifts; optimal IRS phase shifts for maximizing the received signal strength of the combined CE-to-tag link only; and optimal IRS phase shifts for maximizing the received signal strength of the combined tag-to-reader link only. It is evident that notable power reductions are realized at all tag locations compared to the non-IRS-aided system, when the IRS phase shifts from the proposed solution are used; and the reductions are maximized when the tag is close to the IRS. For the two benchmark schemes that solely phase-align either the $C$-$I$-$T$ link or the $T$-$I$-$R$ link, they perform close to the MM solution when the tag is closer to the reader or the CE, respectively. This is due to the far-field nature of the respective links providing larger gains, given the tag's location. However, when the tag is roughly halfway between the CE and reader, the proposed MM solution outperforms both benchmarks, as it selectively phase-aligns each reflector with the more favorable of the $C$-$I$-$T$ and $T$-$I$-$R$ links. Unlike the non-IRS benchmark, the minimized transmit power is not symmetric with tag location, suggesting that the location of the IRS also influences the extent of CE power consumption.

\begin{figure}[!t]
\centering
\includegraphics[width=0.485\textwidth]{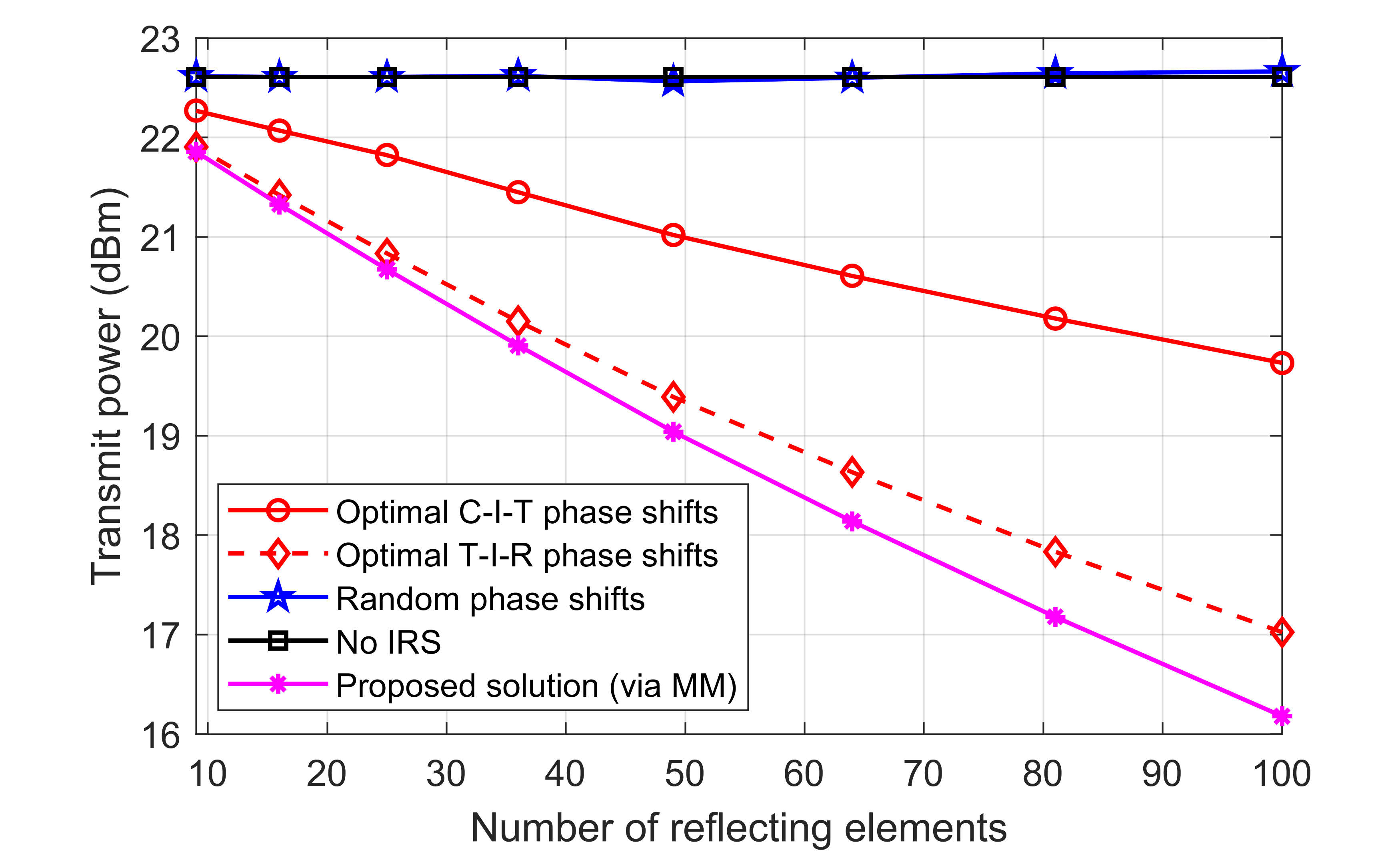}
\caption{Minimized transmit power vs. number of IRS elements.}
\label{fig_3}
\end{figure}

Fig.~3 plots the CE transmit power as a function of the number of IRS elements. The center of the IRS is located at $[20, 20]$ and the tag is located at $[20, 0]$. Compared to the non-IRS benchmark, the CE transmit power scales roughly inversely with the number of IRS elements for the range of $N$ shown. Given that passive devices such as BackCom rely completely on external powering signals for their communication, the reduction of the transmit power consumption is significant. For example, $3.5$ dB power reduction is achieved with a moderately small-size IRS with $N = 49$ while $6.5$ dB is conserved using an IRS with $N = 100$.

\begin{figure}[!t]
\centering
\includegraphics[width=0.485\textwidth]{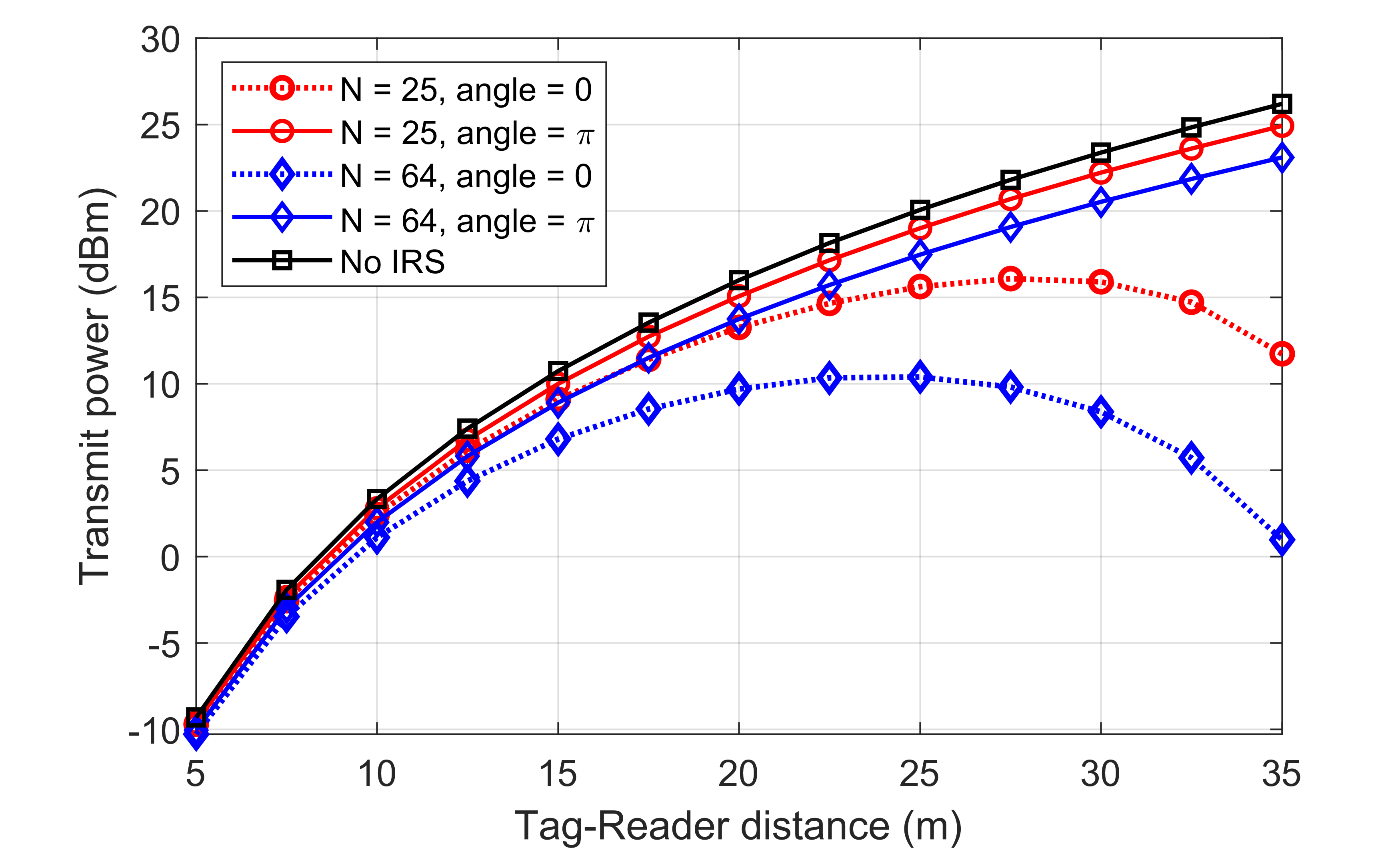}
\caption{Minimized transmit power for a monostatic reader.}
\label{fig_4}
\end{figure}

Fig.~4 reveals the transmit power reductions for a monostatic system with a single-antenna reader as a function of the tag-reader distance, with the center of the IRS located at $[40, 0]$ and oriented towards $[$-$1 \ 0]^{T}$. The angle of the tag relative to the reader-IRS link is varied from $0$ to $\pi$. Similar levels of power reduction are observed, with the largest reduction occurring when the tag is between the reader and IRS and in line with both nodes. One may also observe that while higher transmit power is incurred as the tag-reader distance increases, the extent of power reduction also improves compared to the non-IRS benchmark. It is expected that further range increases can be realized in the case where the reader utilizes multiple antennas to perform transmit and receive beamforming.

\section{Conclusion}
In this paper, a novel IRS-aided bistatic BackCom system was introduced, where the backscatter transmission from the BackCom device is assisted by the IRS. Accounting for multiple additional paths for reflected signals, the transmit power minimization problem at the CE was studied, through the use of the MM algorithm and SDR technique. It was shown that the addition of an IRS, even of moderately small size, is able to considerably reduce the transmit power at the CE. In addition, the optimization problem was extended to monostatic BackCom systems, resulting in similar reductions in transmit power. As an initial work into IRS-aided BackCom systems, there remains much to be studied, including the extension to multi-user scenarios.


%





\ifCLASSOPTIONcaptionsoff
  \newpage
\fi



\bibliographystyle{ieeetran}
\bibliography{IEEEabrv,irs_ref}
\end{document}